\documentclass[conference]{IEEEtran}
\IEEEoverridecommandlockouts
\usepackage{cite}
\usepackage{amsmath,amssymb,amsfonts}
\usepackage{algorithmic}
\usepackage{mathtools}
\usepackage{graphicx}
\usepackage{textcomp}
\usepackage{hyperref}
\usepackage{dsfont}
\usepackage{physics}
\usepackage{xcolor}
\usepackage{soul}
\def\BibTeX{{\rm B\kern-.05em{\sc i\kern-.025em b}\kern-.08em
    T\kern-.1667em\lower.7ex\hbox{E}\kern-.125emX}}

\begin{document}

\title{Warm-Starting MaxCut Relaxation via Low-Depth\\ Quantum Approximate Optimization Algorithm}

\author{
\IEEEauthorblockN{Bao G. Bach}
\IEEEauthorblockA{\textit{Computer and Information Sciences} \\
\textit{Quantum Science and Engineering} \\
\textit{University of Delaware}\\
Newark DE, USA \\
baobach@udel.edu}
\and
\IEEEauthorblockN{Ilya Safro}
\IEEEauthorblockA{\textit{Computer and Information Sciences} \\
\textit{Physics and Astronomy} \\
\textit{University of Delaware}\\
Newark DE, USA \\
isafro@udel.edu}
\and
\IEEEauthorblockN{Filip B. Maciejewski}
\IEEEauthorblockA{
\textit{USRA Research Institute for }\\ \textit{Advanced Computer Science (RIACS),}\\
Moffett Field CA, USA \\
fmaciejewski@usra.edu}
}

\maketitle

\begin{abstract}
Quantum optimization has attracted growing interest as quantum hardware continues to improve, yet state-of-the-art classical solvers remain a formidable benchmark for practical utility. 
Rather than seeking a fully quantum replacement for classical optimization, we propose a hybrid strategy that uses quantum information to enhance leading classical heuristics. 
Specifically, we introduce a warm-start method based on local correlators obtained from the Quantum Approximate Optimization Algorithm (QAOA), and use this information to initialize the Burer–Monteiro (BM) rank-two relaxation. 
We demonstrate numerically that, compared to a random, multi-start initialization baseline (a standard strategy used for BM), this quantum-informed initialization offers a significant head start, i.e., high-quality solutions with very small number of iterations, for two problem classes --  random Erd\H{o}s R\'{e}nyi graphs with edge density of $10\%$ (ER-10) and fully-connected Sherrington Kirkpatrick (SK) spin glass models, at $n=500$ and $n=1000$ qubits.
At the same time, given enough iterations, the random baseline often eventually catches up and slightly outperforms the warm-start strategy on average, an effect visibly stronger for $n=500$ than for $n=1000$.
The results demonstrate an exploitation/exploration tradeoff of using WS to quickly arrive at very good solutions vs exploring slightly better solutions with a larger iterations budget via a standard strategy.
Our results highlight how low-depth quantum circuits can provide useful structural information for classical optimization and suggest a promising route toward near-term quantum utility through quantum-assisted initialization.\\
Reproducibility: the source code and data will be available at~\cite{quapopt_repo}.
\end{abstract}


\section{INTRODUCTION}
The MaxCut problem is a central problem class in combinatorial optimization \cite{dunning2018works}, with applications ranging from finance \cite{herman2023quantum} to graph partitioning \cite{bulucc2016recent}, and statistical physics \cite{lucas2014ising}. At the same time, it is a demanding benchmark for new optimization paradigms, because both its theoretical and practical classical baselines are already strong. On the one hand, the Goemans-Williamson semidefinite programming (SDP) with hyperplane rounding \cite{goemans1995improved} provides the optimal polynomial-time approximation guarantee for unweighted MaxCut under the Unique Game Conjecture \cite{khot2007optimal}. On the other hand, low-rank nonconvex heuristics for both weighted and unweighted graphs, particularly the rank-two Burer-Monteiro relaxation \cite{burer2002rank}, often deliver highly competitive practical performance \cite{dunning2018works} at a substantially lower computational cost than full SDP. Any claim of near-term quantum utility for MaxCut must therefore be evaluated against a very high classical bar.

This creates a challenge for quantum optimization. Algorithms such as the Quantum Approximate Optimization Algorithm (QAOA) \cite{farhi2014quantum} are often studied as candidate solvers for combinatorial optimization \cite{montanez2025toward} including recent scalability experiments in which small-scale quantum solvers are integrated into classical driving routine to achieve a global solution for large-scale problems \cite{bach2024mlqaoa,bach2025solving,ushijima2021multilevel}, but directly outperforming state-of-the-art classical methods remains difficult, especially at low circuit depth \cite{marwaha2021local, wurtz2021maxcut}. This difficulty is strengthened by the maturity and breadth of modern classical optimization. The strongest classical solvers are rarely single-purpose algorithms; they are often carefully engineered combinations of relaxations, local search, preprocessing, rounding, restarts, and other heuristics designed to perform well across many instance classes. Thanks to decades of progress in combinatorial scientific computing and computational optimization, these methods are scalable, robust, and highly practical. 

One widely used ingredient in such methods is initialization, which can strongly influence the trajectory and final quality of solvers. This is the ingredient we explore in this work: rather than using QAOA as a standalone replacement for a classical solver, we ask whether shallow quantum circuits can provide useful structural information that improves a strong classical heuristic. 
In particular, we use an efficient simulation of the local correlators of depth-one QAOA circuits~\cite{wang2018quantum,ozaeta2022expectation}, in hopes that enough information about the distribution of promising cuts (weighted-MaxCut solutions) can be extracted to Warm-Start classical heuristics.

While most prior warm-starting work has been done in the opposite direction \cite{egger2021warm,tate2023bridging,kulshrestha2022beinit} (i.e., using classical methods to initialize quantum optimization or distributions), a growing body of recent work has explored this broader quantum-assisted viewpoint. Quantum information has been used for relax-and-round procedures, preconditioning of classical optimization landscapes, and warm starts for classical heuristics\cite{dupont2025optimization, finvzgar2023quantum, bravyi2020obstacles, vcepaite2025quantum}. However, externally warm-starting the highly performant rank-two Burer-Monteiro (BM) heuristic for weighted MaxCut problems using QAOA correlators remains comparatively underexplored -- to the best of our knowledge, there is no prior work in this direction. To fill in this gap, we developed a warm-start strategy that uses QAOA correlators to propose good initial points for the BM solver with a preparation cost that is a modest fraction of a full run. 
Numerically, we study whether this QAOA-informed initialization accelerates convergence and improves solution quality relative to standard multi-start random initializations. 
More broadly, our results support the view that near-term quantum utility in optimization may be more realistically achieved through quantum-assisted initialization and preconditioning than through end-to-end quantum optimization alone.

\begin{figure*}
    \centering
    \includegraphics[width=.8\linewidth]{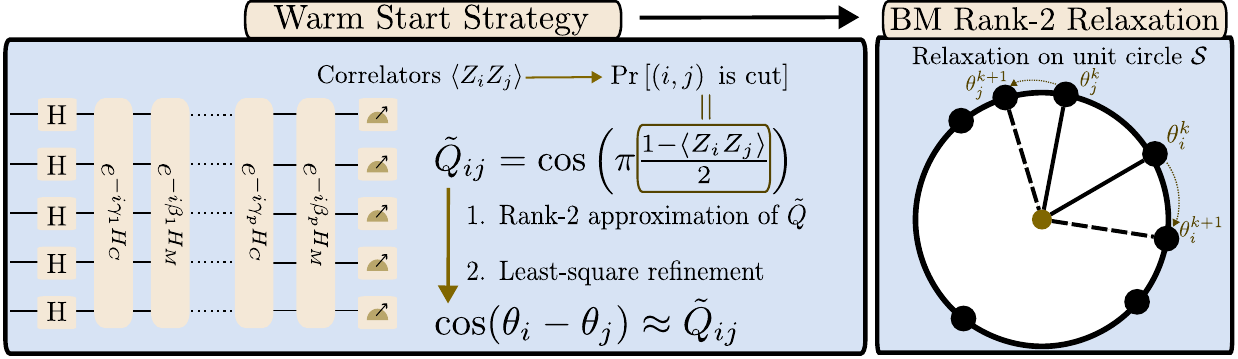}
    \caption{Warm Start Framework for Burer-Monteiro rank-two relaxation using QAOA correlators $\langle Z_{i} Z_{j}\rangle$. 
    The corresponding marginal distributions of the probability of edge $(i,j)$ being cut are then mapped via a Goemans-Williamson-inspired trigonometric function to $\tilde{Q}$.
    We then find the closest rank-2 correlation matrix via projecting onto the two largest positive eigenvalues of $\tilde{Q}$, followed by least-squares error optimization.
    The resulting matrix is then interpreted as encoding the angles $\theta$ that are used as the initial point for Burer-Monteiro rank-two relaxation.
    }
    \label{fig:main_fig}
\end{figure*}

In summary, this paper makes the following contributions:
\begin{enumerate}
    \item We introduce a Warm-Start strategy for the rank-two Burer-Monteiro heuristic for weighted MaxCut, which uses pairwise cut probabilities from QAOA to propose good initial angles for the BM solver.
    This is achieved by a heuristic least-squares error minimization that aims to find the closest rank-2 correlation matrix consistent with local QAOA expected values, which in turn can be used to infer starting angles for BM.
    \item We provide numerical evidence that this quantum-informed initialization, with just depth-1 QAOA, allows Burer-Monteiro to arrive at very good solutions very quickly, providing a significant head-start at a small iteration budget. 
    However, given enough budget, a random multistart strategy can eventually slightly outperform the WS for some instances. Notably, this effect is visibly weaker at the $n = 1000$-qubit instances (larger system).
\end{enumerate}
A schematic illustration of our pipeline is shown in Fig.~\ref{fig:main_fig}.

\section{PRELIMINARIES}
\subsection{Problem Formulations}

We consider the weighted MAXCUT problem, which is NP-complete \cite{garey1974some} and is widely used as a benchmark for QUBO algorithms \cite{dunning2018works} (any $n$-variable QUBO can be mapped to $\left(n+1\right)$-variable MAXCUT\cite{boros1991max}). 
For a graph $G(V, E)$ with $|V| = n$, the MAXCUT problem seeks a partition of the vertex set into two disjoint subsets, $V_1$ and $V_2$, such that the total weight of edges crossing between the two subsets is maximized. 
The graph can be represented by an upper-triangular weight matrix $W \in \mathbb{R}^{n \times n}$ with zero diagonal, yielding the objective
\begin{align}
  \label{eq:MC_spin_formulation}
  \max_s \sum_{i=0}^{n-1} \sum_{j=i}^{n-1} \frac{W_{i,j}}{2}
  \left(1-s_is_j\right)
  ,  
\end{align}
where we have classical spin solution $s \in \left\{-1,1\right\}^{n}$.

MAXCUT can be formulated as an Ising model without external fields \cite{barahona1988application}, by promoting classical spins $s_i$ to Pauli $Z_i$ operators and identifying weights $W_{ij} \rightarrow J_{i,j}$
\begin{align}\label{eq:ising_hamiltonian}
    H_{C} = \sum_{i,j}J_{i,j}Z_{i}Z_{j} \ .
\end{align}
The above formulation is used in quantum optimization methods to solve weighted MAXCUT problems.

\subsection{Best Polynomial-time Algorithm For Max Cut}
The Goemans--Williamson (GW) algorithm \cite{goemans1995improved} is the standard polynomial-time approximation algorithm for non-negative MAXCUT and, assuming the Unique Games Conjecture, it is optimal among all polynomial-time approximation algorithms \cite{khot2007optimal}. 
To introduce the GW algorithm, consider a positive-semidefinite (PSD) matrix $X = ss^T$ defined for a candidate solution $s$. 
Then $X_{i,j} = s_i s_j$, $X \succeq 0$, $\operatorname{diag}(X)=\mathbf{1}$, and $\operatorname{rank}(X) = 1$. The Goemans--Williamson relaxation is obtained by dropping the nonconvex rank-one constraint, which yields the following SDP
\begin{equation}
    \label{eq:SDP}
   \begin{aligned}
    \max_{X}& \sum_{i=0}^{n-1}\sum_{j=i}^{n-1} \frac{W_{i,j}}{2}(1 - X_{i,j}) \\
    \text{s.t. }&
    X \succeq 0,\;
    \operatorname{diag}(X)=\mathbf{1}_{n}.
\end{aligned} 
\end{equation}
Since $X\succeq 0$ with $1$s on the diagonal, it can be interpreted as a Gram matrix of some unit vectors $\left\{v_{1}, \dots, v_{n}\right\}$, and the relaxed objective can be rewritten as
\begin{align}\label{eq:vector_relaxation}
    \max \sum_{i=0}^{n-1}\sum_{j=i}^{n-1} \frac{W_{i,j}}{2}\bigl(1 - \langle v_i,v_j\rangle\bigr).
\end{align}
After solving this SDP, the algorithm rounds the vector solution to a discrete cut by drawing a random hyperplane through the origin and assigning vertices according to the signs of their projections onto the hyperplane. Goemans and Williamson proved that this procedure achieves an expected approximation ratio of $\alpha_{\mathrm{GW}} \approx 0.87856$ with respect to the optimal cut value \cite{goemans1995improved} for non-negative weights.
While this ratio is proved only for non-negative weights, the GW algorithm and GW-inspired heuristics (see below) remain excellent practical solvers also for general-weight (including negative and positive) problems considered in this work.

\subsection{Competitive Heuristic For MaxCut}
Based on the established benchmark from \cite{dunning2018works}, where multiple solvers are tested over a variety of graph instances, the Burer-Monteiro rank-$2$ relaxation heuristic \cite{burer2002rank} gives the best performance over multiple criteria. The heuristic adapts the SDP formulation of Goemans-Williamson and proposes a lower-rank nonconvex alternative to obtain high-quality cuts more efficiently in practice. 
This is achieved by adding a rank-2 constraint to Eq.~\eqref{eq:SDP}, which can be ensured 
by introducing the angle vector $\theta$ and parameterizing 
\begin{equation}
    X(\theta) = C(\theta) C^{T}(\theta)+S(\theta)S^{T}(\theta) ,
\end{equation}
where $C_i(\theta) = \cos\left(\theta_i\right)$, and $S_i(\theta) = \sin\left(\theta_i\right)$, each vertex being associated with angle $\theta_i$ or, equivalently, a point on a unit circle.
Then, the relaxation becomes an unconstrained but nonconvex minimization of the $n$-variable function
\begin{equation}\label{eq:cost_func}
    f(\theta) = \sum_{i,j}\frac{W_{i,j}}{2} \cos(\theta_{i}-\theta_{j}) .
\end{equation}
Note that the above is related to Eq.~\eqref{eq:vector_relaxation} via identity $\langle v_i,v_j\rangle = \cos\left(\theta_i-\theta_j\right)$ for $2$-dimensional vectors.
The heuristic first applies a local continuous optimization method to minimize $f(\theta)$ and obtain locally optimal angles.
It then binarizes the angles by running a diameter through the circle, and assigning the binary values based on which side of the diameter the corresponding point lies. 
The final position of the diameter corresponds to the best-cut value, which can be efficiently found by sweeping a diameter through the circle, and choosing the best-cut configuration via brute force. For each $\alpha\in[0,\pi)$, one assigns $x_i=+1$ if $\theta_i\in[\alpha,\alpha+\pi)$ and $x_i=-1$ otherwise, and selects the best cut produced by this sweep.
The algorithm is then restarted from a small random perturbation of the angular representation of the current cut, to potentially escape the local minimum. 
In practice, this procedure is often supplemented by a 1- and 2-local sweep (bit flips) on the binary solution, and we use the 2-local sweep strategy in the implementation as in \cite{dunning2018works} in this work.

\subsection{Quantum Approximate Optimization Algorithm}
The Quantum Approximate Optimization Algorithm (QAOA) \cite{farhi2014quantum} is a hybrid quantum-classical algorithm designed for combinatorial optimization problems. 
Given an Ising cost Hamiltonian (Eq.~\eqref{eq:ising_hamiltonian}), depth-$p$ QAOA ansatz consists of alternating layers of Phase Separator unitary $U_{PS}(\gamma) = e^{-i\gamma H_C}$ and Mixer unitary $U_{M}(\beta) = e^{-i\beta H_M}$, parametrized by length-$p$ vectors of angles $\gamma$ and $\beta$; applied to initial state $\ket{+}^{\otimes n}$.
Here, $H_{M} = \sum_{i}X_i$.
The QAOA subroutine involves preparing the parametrized state
\begin{equation}
    \label{eq: QAOA_state}
    \ket{\gamma,\beta} = U_{M}(\beta_p)U_{P}(\gamma_p)\dots U_{M}(\beta_1)U_{P}(\gamma_1)\ket{+}^{\otimes n} \ ,
\end{equation}
and optimizing the expected value of the energy
\[
\langle H_f \rangle = \bra{\gamma,\beta}H_f\ket{\gamma,\beta}.
\]
See Ref.~\cite{gemeinhardt2023quantum} for an overview of multiple variants generalizing the above vanilla QAOA.

\section{Related Works}

\subsection{Warm Start Rank-k Relaxation Heuristic}
To our knowledge, there is little prior work devoted specifically to warm-starting the rank-two Burer-Monteiro heuristic for MAXCUT from an external solver. The original rank-two method of Burer, Monteiro, and Zhang already incorporates an internal restart mechanism: after optimizing the angular rank-two relaxation and rounding it to a cut, the algorithm perturbs the angular representation of that cut and resolves the continuous problem \cite{burer2002rank}.  

\subsection{Warm Start Classical Solver using Quantum Information}
There is a small but growing body of work on using quantum-generated information to accelerate classical optimization routines. 
One important example is the quantum relax-and-round (QRR) framework of Dupont and Sundar~\cite{dupont2024extending}, which uses the $\langle Z_iZ_j \rangle$ correlations to find binary solutions via eigensolving the correlation matrix and rounding the leading eigenvector. 
Closely related work on \textit{quantum preconditioning} demonstrates that shallow-QAOA correlation data can be used to transform the original instance into a related one on which strong classical solvers, including simulated annealing and the Burer-Monteiro heuristic, converge more rapidly on Max-Cut-type benchmarks \cite{dupont2025optimization}. We note that quantum preconditioning \emph{does not} Warm-Start BM; instead, it replaces the problem being solved by BM with QAOA-obtained correlators matrix.
A related recursive approach based on local expected values was presented in Ref.~\cite{finvzgar2023quantum}, building on the recursive/iterative QAOA framework~\cite{bravyi2020obstacles, brady2024iterative}.
In Ref.~\cite{vcepaite2025quantum}, \v{C}epait\.e \textit{et al.} propose a \textit{quantum-enhanced optimization by warm starts} approach in which measured samples from quantum circuits are used explicitly as an initial solution for classical heuristics (tabu search~\cite{glover1990tabu}) on combinatorial optimization problems such as Max-Cut and Maximum Independent Set.

Taken together, these works suggest that using quantum information not necessarily to replace classical optimization, but rather to initialize, precondition, or guide classical solvers, is becoming a distinct and increasingly active research direction.
Notably, low-rank Burer-Monteiro solutions for MAXCUT have been historically used also in the opposite direction, namely as classical warm-starts for the quantum optimization \cite{tate2023bridging, egger2021warm}. Overall, while restart and continuation ideas are clearly present in the literature, externally warm-starting the rank-two MAXCUT heuristic itself appears to be comparatively underexplored.

\section{Methodology}
We investigate how to warm-start the best-known pratical heuristic, the Burer-Monteiro rank-two relaxation for MaxCut, using local correlators of the depth-one Quantum Approximate Optimization Algorithm.
\subsection{2-Local Correlators And MaxCut}
Consider QAOA state $\ket{\psi(\gamma, \beta)}$ from Eq.~\eqref{eq: QAOA_state}. 
While drawing samples from the probability distribution corresponding to that state is believed to be computationally hard in general~\cite{farhi2016quantum}, it turns out that for $p=1$ QAOA and $2$-local Hamiltonians, one can derive closed-form expressions for the expected values of local correlators $\langle{Z_{i}Z_{j}}\rangle$ that are efficient to evaluate, see, e.g., Refs.~\cite{wang2018quantum, ozaeta2022expectation}.
This means that we can efficiently optimize the expected value $\langle H_{C}\rangle$ in classical simulations, and access the expected values of the correlators on the optimized state. 
Consider now solution $s$ and edge $e=\left(i,j\right)$. 
If $s_i\neq s_j$, then $e$ is cut.
For each edge, we can calculate the corresponding marginal probability distribution as 
\begin{equation}\label{eq:pij_qaoa}
    \begin{aligned}
        \mathrm{Pr}\left[\left(i,j\right)\ \mathrm{is\ cut}\right] \coloneqq  p_{i,j} =  \frac{1 - \langle Z_{i}Z_{j}\rangle}{2}
    \end{aligned}
\end{equation}

\subsection{Warm Start using QAOA Correlators}

Recall that in BM rank-two relaxation, each vertex $i \in V$ is associated with a point on a unit circle, $\theta_i \in \left[0, 2\pi\right]$, and the optimized cost function (Eq.~\eqref{eq:cost_func}) involves $\cos(\theta_i-\theta_j)$ terms. 
Our goal is to find $\theta = (\theta_{1}, \dots, \theta_{n})$ for which the correlation structure is consistent with QAOA correlators. 
To this aim, we recall that the hyperplane rounding in the GW algorithm assigns the edge-cut probability via $\mathrm{Pr}_{e}(\bar{\theta}) =  \arccos(\langle v_i, v_j \rangle)/\pi = \abs{\theta_i-\theta_j}/\pi $ \cite{goemans1995improved}.
We thus would like to find such $\theta$ that
\begin{align}
    \label{eq:hr_approx}
    \cos(\theta_{i} - \theta_{j}) \approx \cos \left(\pi\ p_{i,j}\right)\ .
\end{align}
To this aim, we define the matrix 
\begin{align}
    \tilde{Q}_{i,j} \coloneqq  \cos \left(\pi\ p_{i,j}\right) = \sin\left(\frac{\pi}{2}\left<Z_iZ_j\right>\right) , 
\end{align}
and the corresponding nonlinear least-squares minimization problem as
\begin{align}
    \label{eq:non_linear_lsq}
    \min_{\theta \in[0, 2\pi)^n} F(\theta) = \sum_{ij \in E}(\cos(\theta_i - \theta_j) - \tilde{Q}_{i,j})^2 , 
\end{align}
with direct calculation yielding gradients
\begin{align}
    \frac{\partial F}{\partial \theta_i} = \sum_{k \sim i} -2\sin (\theta_{i}-\theta_{k})(\cos(\theta_{i}-\theta_{k}) - \tilde{Q}_{i,k})
\end{align}

In other words, we are trying to find the closest (to the $\tilde{Q}$) rank-2 correlation matrix.
Noting the similarity of Eq.~\eqref{eq:non_linear_lsq} to the original relaxed cost function of Eq.~\eqref{eq:cost_func}, we can use the same heuristic to solve it.
In particular, we use the Armijo gradient line search, as implemented in the \texttt{MQLib} library for standard BM~\cite{dunning2018works}
We remark that with this choice, the numerical computation required to find Warm-Started angles is of similar complexity as a single iteration (corresponding to one restart) of the standard BM algorithm -- remembering that we still require QAOA results as an input.

While the minimization in Eq.~\eqref{eq:non_linear_lsq} can be initialized from a random point, we found it beneficial to use a rank-2 approximation of the $\tilde{Q}_{i,j}$ matrix to derive an initial-point ansatz. 
To this aim, we perform eigendecomposition of $\tilde{Q}$ and use the two eigenvectors $v_{1}, v_{2}$ corresponding to the two largest positive eigenvalues $\lambda_1, \lambda_{2}$ to construct a matrix that is then interpreted as encoding initial angles via $\theta_{i, \tilde{Q}} = \mathrm{atan2}(\sqrt{\lambda_{2}}v_{i, 2}, \sqrt{\lambda_{1}}v_{i, 1})$. A common angular shift is subsequently applied to fix the global rotational gauge.

\section{Numerical Experiments}

\subsection{Methodology}
We numerically test two classes of problems. 
The first class, ER-10, consists of Erdős-Renyi random graphs, where each edge is added to the graph with probability $10\%$ from the pool of all possible edges. 
The second class, SK, corresponds to the Sherrington-Kirkpatrick spin glass model, where the graphs are fully connected.
The edge weights for both classes are sampled from a normal distribution with mean $0.0$ and standard deviation $1$.
Both classes are tested for $n=500$ and $n=1000$ nodes.
For ER-10 at both sizes, and SK at $n=500$, we generate $500$ random problem instances, and $300$ for SK at $n=1000$.
For each instance, we implement the Burer-Monteiro relaxation that is capped at $1000$ iterations, where each iteration involves adding a random perturbation to the previous iteration's post-2-local-sweep solution (the current cut)~\cite{burer2002rank,dunning2018works} -- and restarting the solver if $11$ consecutive iterations did not improve the energy. 
We implement the following three restart strategies:
\begin{enumerate}
    \item Warm-Start -- the initial point of BM is obtained from QAOA correlators via the procedure described previously, and solver restarts go back to the same initial point (with new random perturbations).
    \item Random MultiStart -- the reference baseline, where the initial point is random, and solver restarts correspond to new random initialization.
    \item RandomLocal -- the WS-matched-policy reference, where the solver's initial point is random, and solver restarts go back to that fixed random point.
\end{enumerate}
For non-fully-connected ER-10 instances, we additionally test two sub-strategies, where in Eq.~\eqref{eq:non_linear_lsq} we either consider only edges present in the interaction graph (WS-GraphOnly strategy), or all $\binom{n}{2}$ two-body terms (WS-AllPairs strategy). Note that, WS-AllPairs eigendecomposes the full $\tilde{Q}$ while WS-GraphOnly first sets non-edge entries of $\tilde{Q}$ to zero and then performs the eigen-decomposition. Thus, the mask affects both the spectral initial point and the least-squares refinement.

For each instance, each strategy is run with $3$ different initialization seeds, and we take the mean as the point estimate for a fixed instance. All statistical functions presented in this work are applied over an ensemble of Hamiltonian instances (and for each instance the datapoints are the averages over those $3$ seeds).

To compare solvers, we use three figures of merit
\begin{enumerate}
    \item Best energy found by the solver.
    \item The paired difference between the WS solver best-found energy and the Random Multistart baseline. 
    \item WinRate of the solver, defined as the estimated probability that the solver performs at least as well as the baseline.
\end{enumerate}
The above are plotted as functions of the iteration budget. 
The figures of merit are taken as the mean over the relevant Hamiltonian ensemble.
We use cluster bootstrapping with $s=10^4$ resamples to estimate the plotted $95\%$ confidence intervals on the mean estimators.

\subsection{Results}

\begin{figure*}[t!]
\centering
\includegraphics[width=0.45\textwidth]{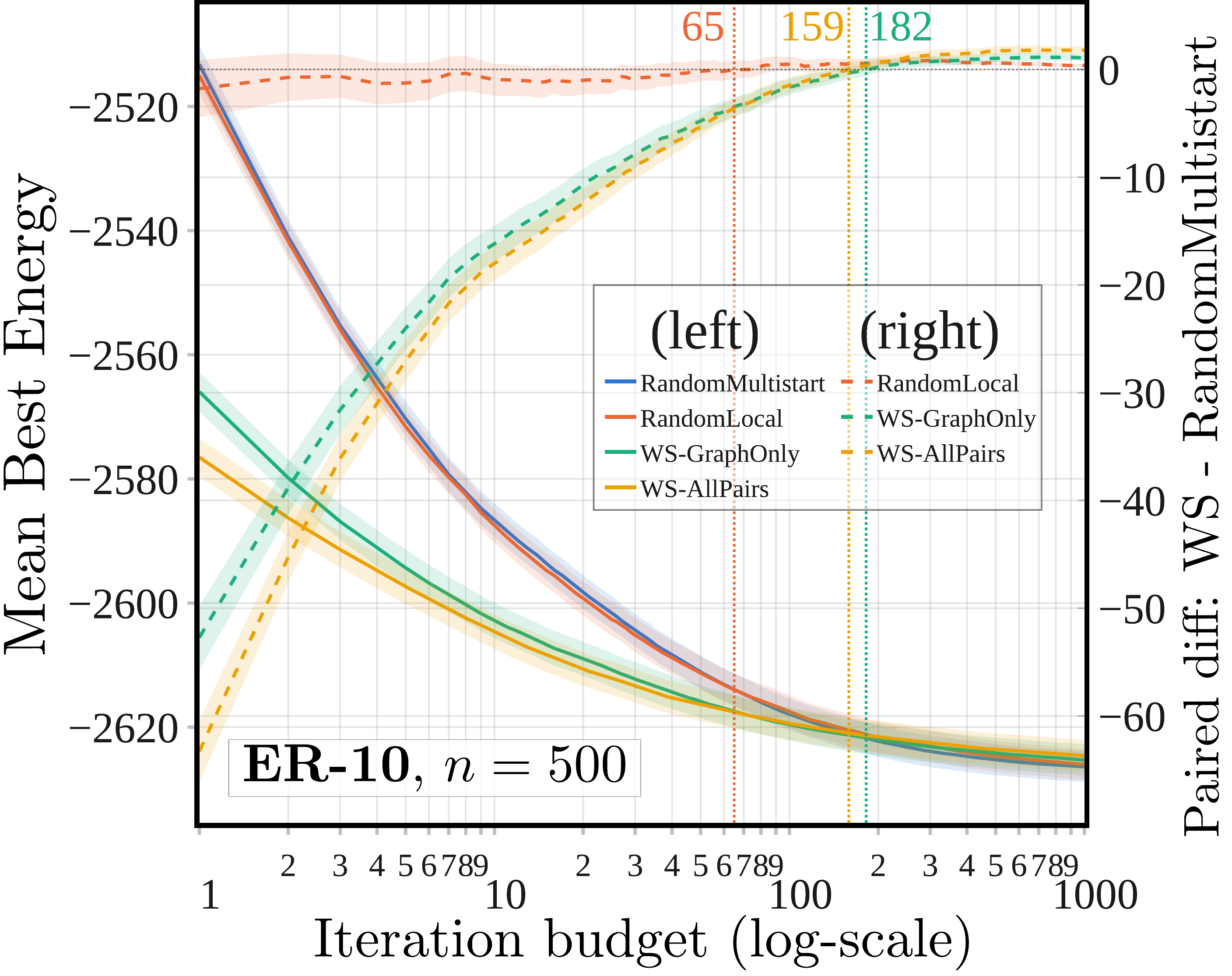}
\quad
\includegraphics[width=0.45\textwidth]{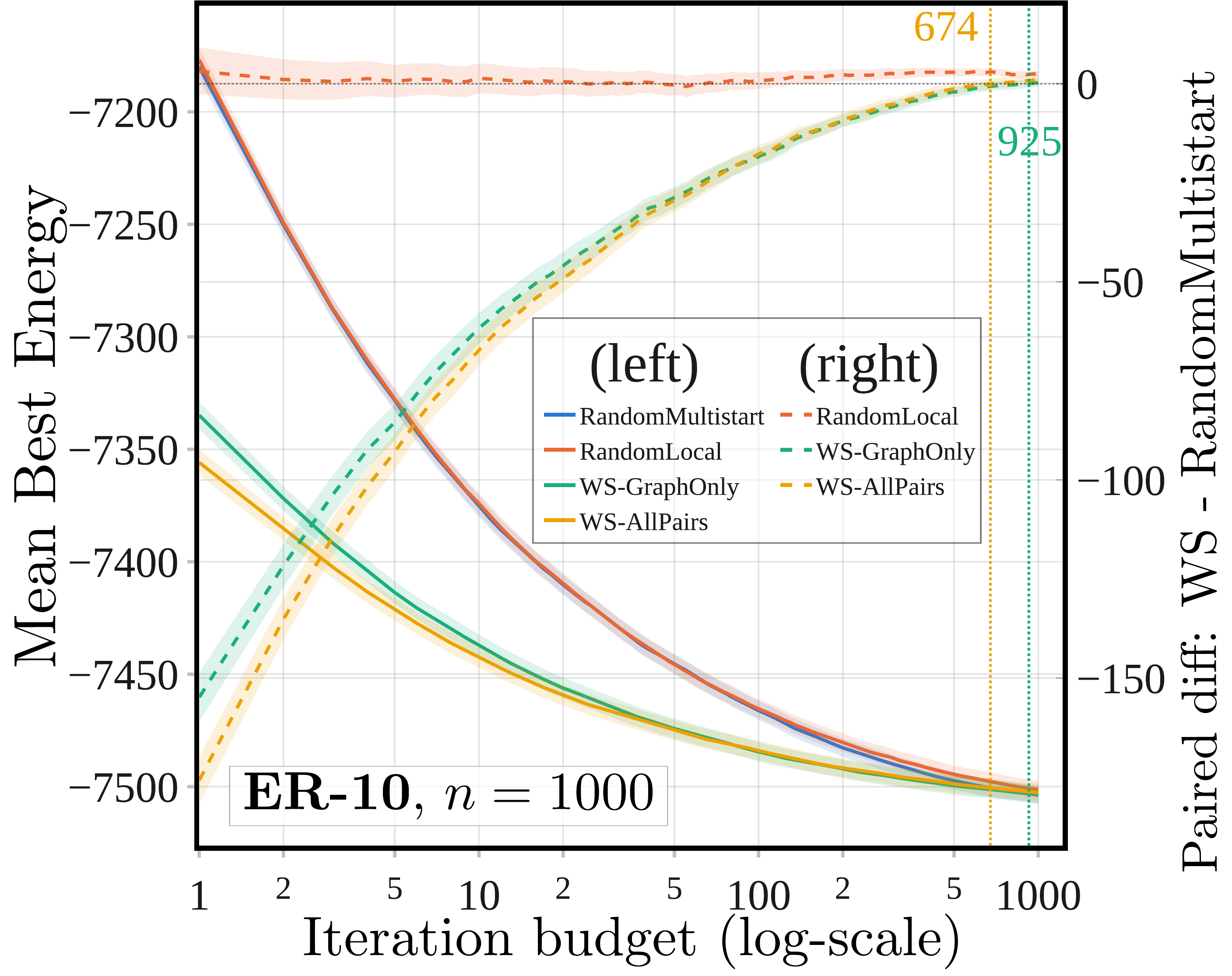} \\ 
\includegraphics[width=0.45\textwidth]{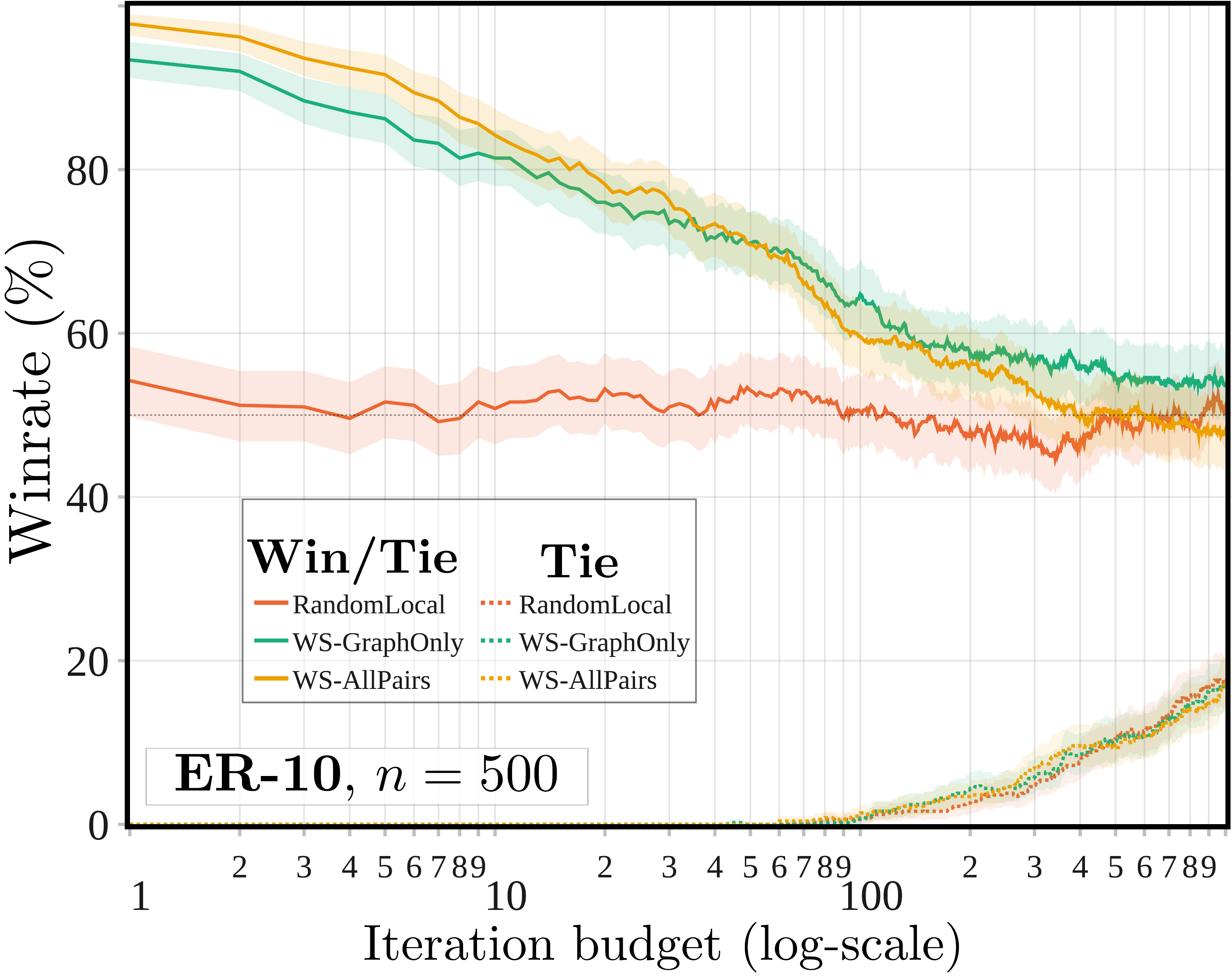}
\quad
\includegraphics[width=0.45\textwidth]{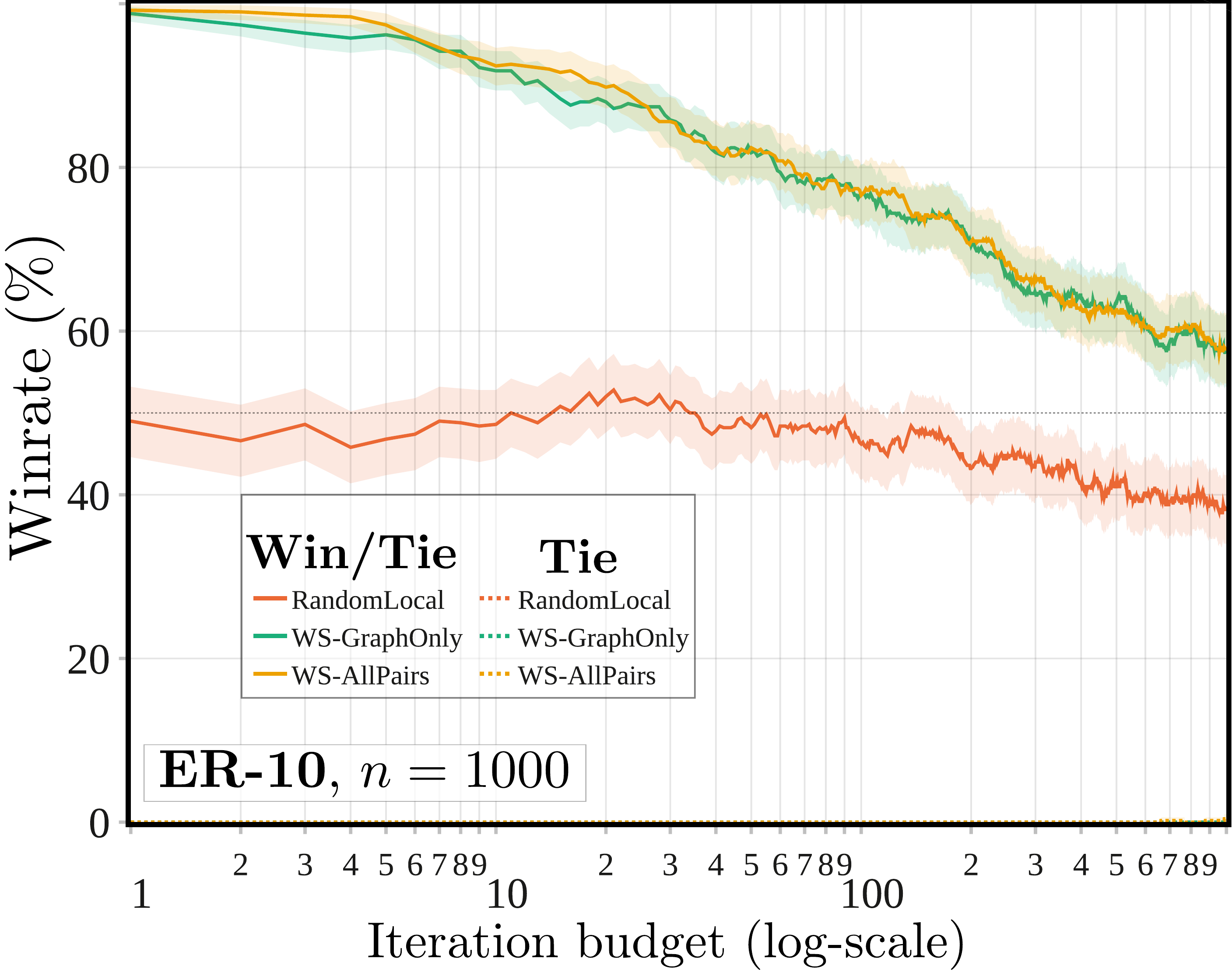}
\caption{\label{fig:ER_results}
The top plots show the best-found energy (left y-axis, solid lines) and paired energy differences (right y-axis, dashed lines) between the given solver (WS with QAOA or reference RandomLocal) and the Random MultiStart baseline as a function of iteration budget, for the ER-10 ensemble with $n=500$ (left column) and $n=1000$ (right column).
The dotted vertical lines (with annotated iterations) correspond to the crossover point where mean stops being ahead of the baseline.
The bottom plots show the WinRate -- the estimate of the probability that the given solver performs at least as well as the baseline at fixed iteration budget (solid lines); this is augmented with relevant tie probability (dotted lines).
The indicated confidence intervals (shadows) are $95\%$ bootstraps with $s=10^4$ resamples over Hamiltonian instances and taking the mean  (not the spread of individual instances). 
}
\end{figure*}

We start by presenting the results for the ER ensemble for both system sizes $n=500, 1000$ in Fig.~\ref{fig:ER_results}.

We observe that the QAOA-based WS strategy starts in a significantly better region (lower energies and paired differences, higher win rate).
In more detail, at $n=500$ the MultiStart baseline crosses the performance of WS at around $159$ iterations for WS-AllPairs, and $182$ iterations for WS-GraphOnly. At $n= 1000$, the corresponding crossing points occur at $674$ iterations for WS-AllPairs and $925$ iterations for WS-GraphOnly. Notably, while for $n=500$ the WS win rate eventually reaches around $50\%$ for the maximal budget of $1000$ iterations, for $n=1000$ it ends closer to $60\%$.
Interestingly, the GraphOnly and AllPairs strategies for ER-10 differ visibly -- the AllPairs seems to start in a better region initially, but GraphOnly eventually catches up and ends up being a better strategy for higher iteration budgets.

\begin{figure*}[t!]
\centering
\includegraphics[width=0.48\textwidth]{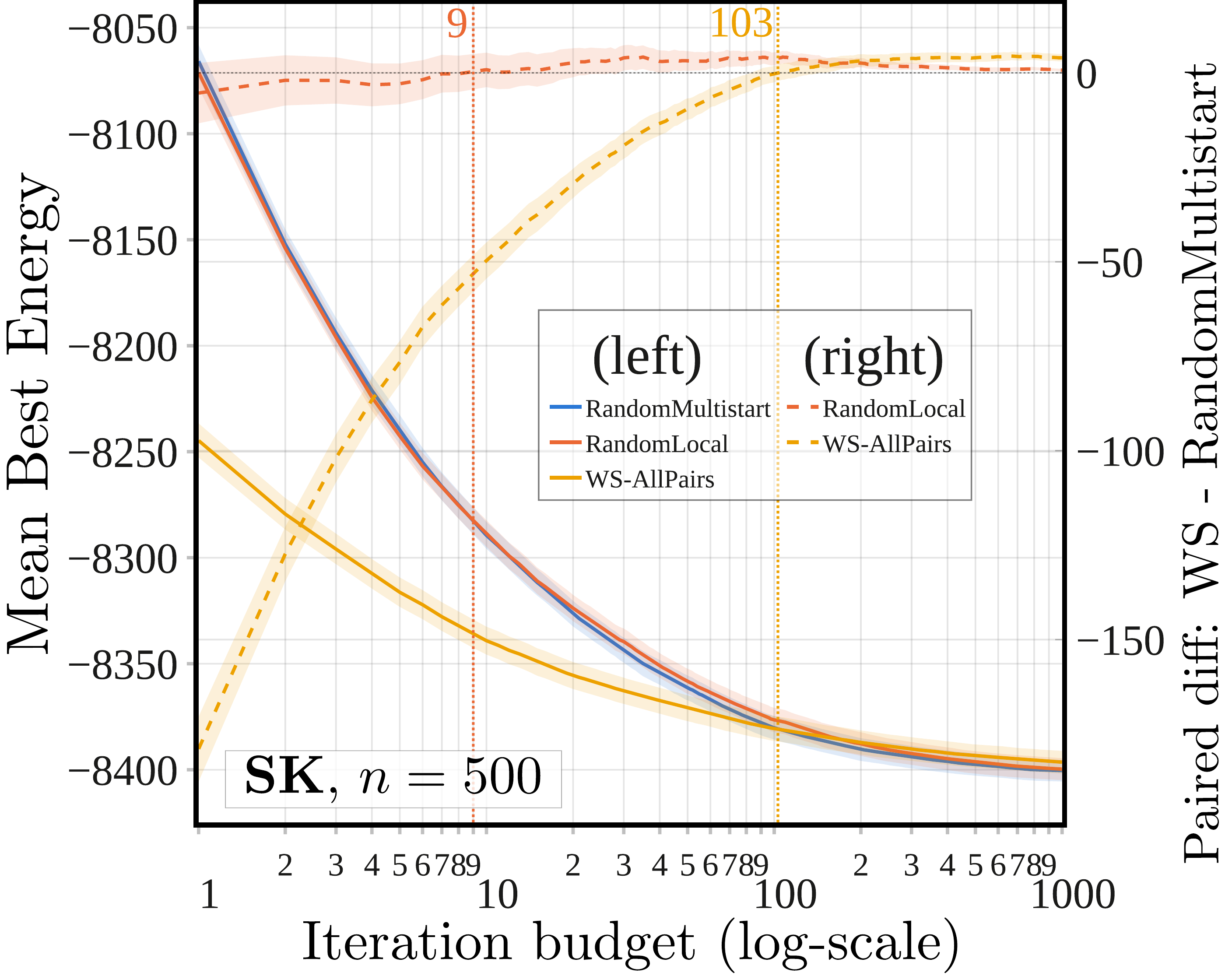}
\quad
\includegraphics[width=0.48\textwidth]{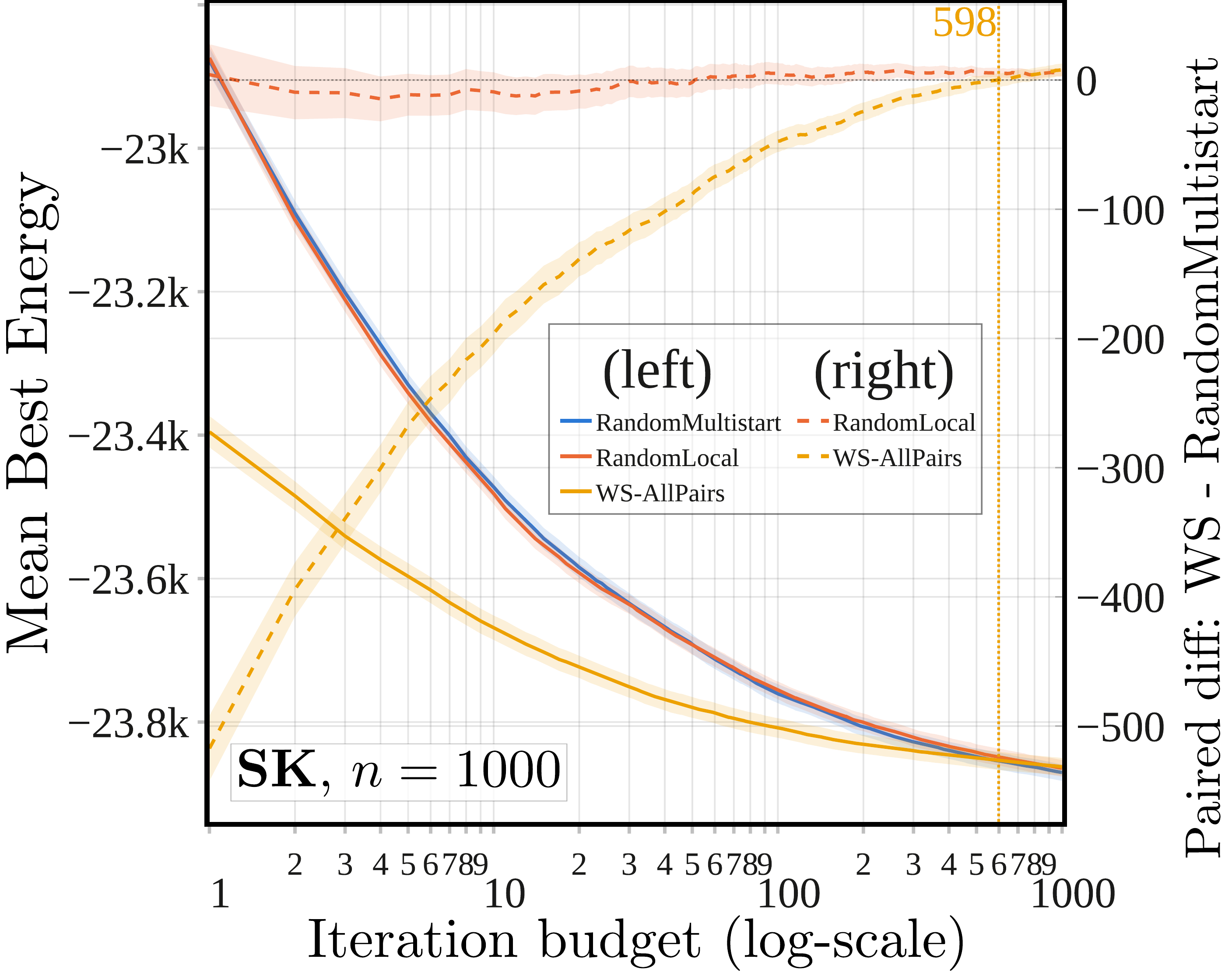} \\ 
\includegraphics[width=0.48\textwidth]{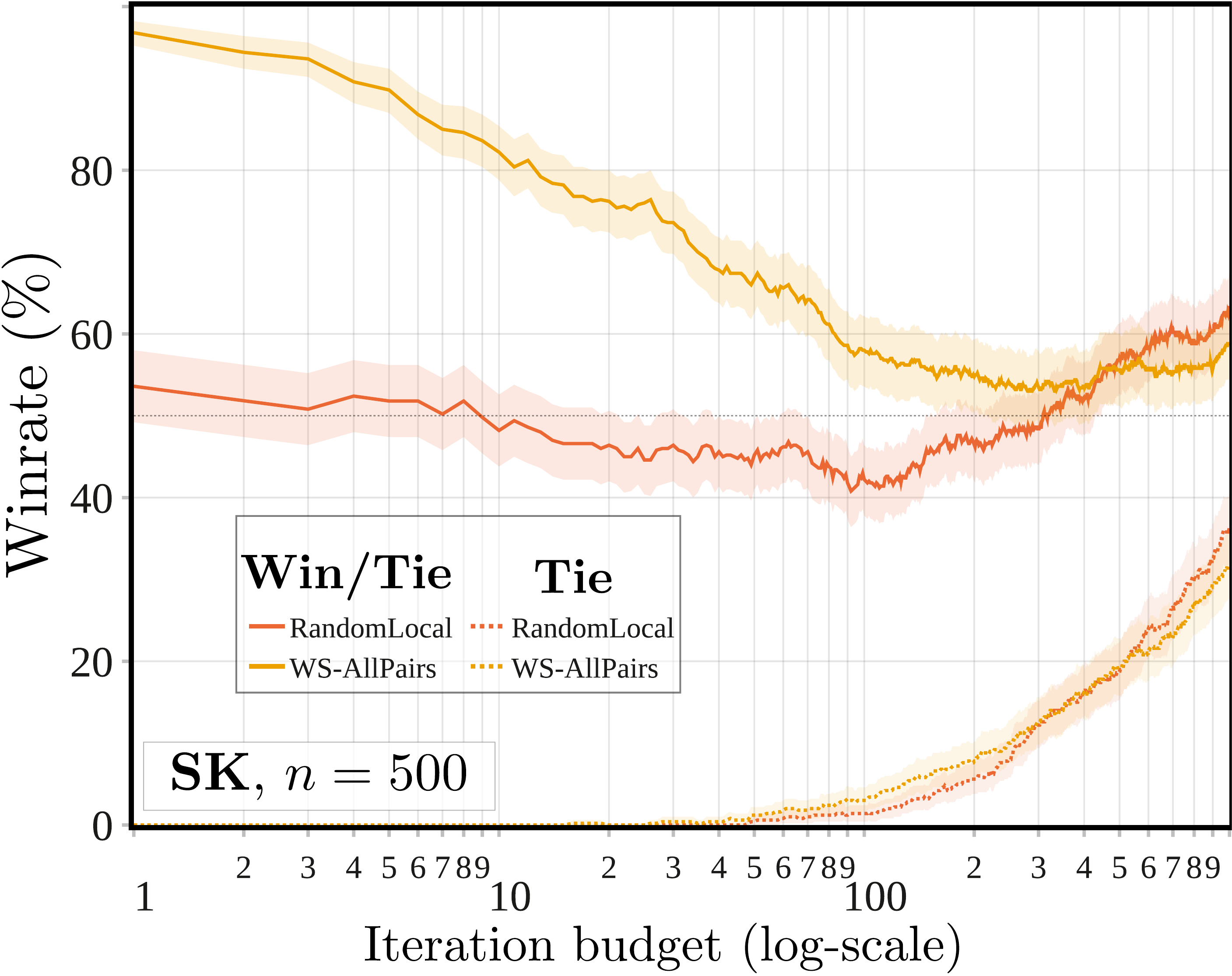}
\quad
\includegraphics[width=0.48\textwidth]{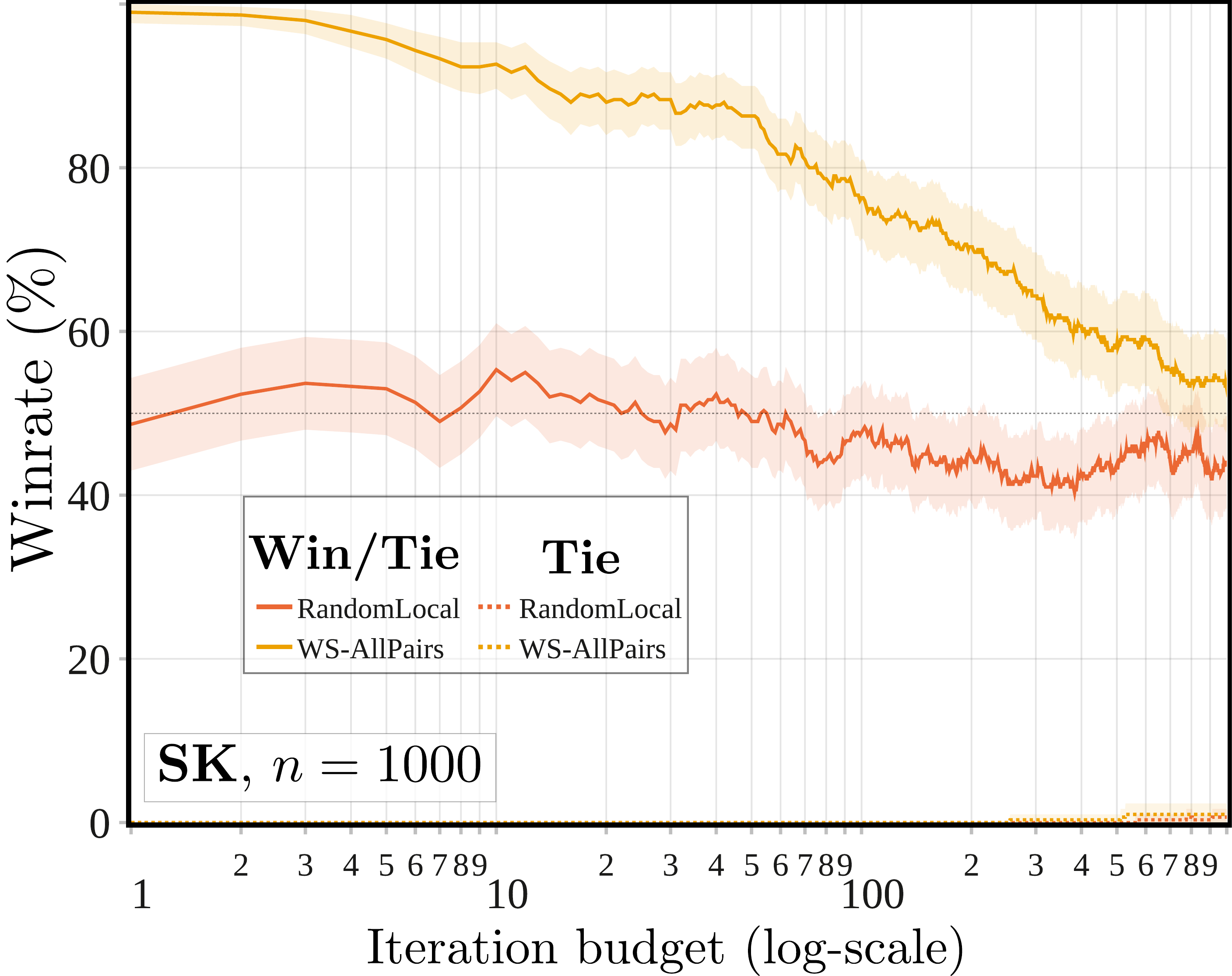}
\caption{\label{fig:SK_results}
Numerical results between different approaches with representation as in Figure.~\ref{fig:ER_results}, for SK instances.
}
\end{figure*}

The results for the SK model are presented in Fig.~\ref{fig:SK_results}, where we observe that the Warm-Start strategy performs generally worse than for ER-10 -- the crossover for $n=500$ appears already around iteration $\approx 100$, and for $n=1000$ for $\approx 600$.

For both classes, the WS for $n=1000$ seems to be generally more effective than for $n=500$, indicating potentially favorable scaling of using the proposed strategy -- to be investigated in future work.
We also note that there is a large number of ties for $n=500$, whereas there are none or almost none for $n=1000$, which means that whenever WS wins, it achieves a better energy than Random Multistart baseline.

\section{Discussion}
We introduced a QAOA-informed warm start for the rank-two Burer–Monteiro MaxCut heuristic by mapping depth-one QAOA pair correlations to an initial angular embedding. Across the ER-10 and SK ensembles, the resulting initialization provides a clear advantage at small iteration budgets, allowing BM to reach high-quality solutions substantially faster than random initialization. The RandomLocal control indicates that this improvement is attributable to the information contained in the QAOA-derived seeds. At larger budgets, however, Random MultiStart often catches up and slightly outperforms the warm-start strategy because independent restarts explore a broader landscape. The results, therefore, reveal an exploitation–exploration tradeoff: the QAOA-informed strategy improves early time-to-quality, whereas random multistart can obtain marginally better solutions when given sufficient search budget. These results demonstrate an improvement in the BM refinement stage rather than an end-to-end quantum computational advantage, since the depth-one correlators are evaluated classically and the reported budgets exclude QAOA optimization and warm-start construction. Nevertheless, they show that QAOA-derived correlations can provide useful structural information for initializing a strong classical solver

\bibliographystyle{unsrt}
\bibliography{bibliography,ilya-biblio}

\section*{Acknowledgment}

 B.G.B. and I.S. acknowledge support under the NSF award \#2444042. F.B.M. acknowledges support under the NSF award \#2329097. We are grateful to Stuart Hadfield, Davide Venturelli, Joao Prioli, and José Carlos Hernández Azucena for useful discussions throughout this project.

\end{document}